\documentstyle[11pt]{article}

\newtheorem{theorem}{Theorem}[section]

\newtheorem{proposition}[theorem]{Proposition}
\newtheorem{corollary}[theorem]{Corollary}

\def\qed{\hfill $\Box$\medskip}

\def\IR{{\bf R}}

\begin{document}
\baselineskip 14.5pt
\begin{center}
{\large{\bf\sc Decomposition of unitary matrices and quantum gates}}

\bigskip
{\small\sc Chi-Kwong Li\footnote{Corresponding author.}, Rebecca Roberts}

\smallskip
{\small\sl
Department of Mathematics, College of William and Mary, Williamsburg, VA 23187, USA.}

(E-mail: ckli@math.wm.edu, rlroberts@email.wm.edu)

\medskip
{\small\sc Xiaoyan Yin}

\smallskip
{\small\sl Department of Mathematics, Xidian University,
Xi'an, Shaanxi, 710071, China. }

(E-mail: yinxiaoyan0105@126.com)

\end{center}

\medskip
\noindent
{\bf Abstract}

{\small A general scheme is presented to decompose a $d$-by-$d$ unitary matrix
as the product of two-level unitary matrices with additional structure
and prescribed determinants. In particular, the decomposition can be
done by using two-level matrices in $d-1$ classes, where each class is
isomorphic to the group of $2\times 2$ unitary matrices. The proposed scheme
is easy to apply, and useful in treating problems with the additional
structural restrictions.  A Matlab program is written to implement the scheme,
and the result is used to deduce the fact
that every quantum gate acting on $n$-qubit registers can be
expressed as no more than $2^{n-1}(2^n-1)$ fully controlled single-qubit
gates chosen from $2^n-1$ classes, where the quantum gates in each class
share the same $n-1$ control qubits. Moreover, it is shown that 
it is easy to adjust the proposed decomposition scheme
to take advantage of additional structure evolving in the process.}


\section{Introduction}

Matrix factorization is an important tool in matrix theory and its applications.
For example, see the general references \cite{H,GV,HJ}, and some recent papers
\cite{Bet,F,St,Vet}  and the references therein on special topics.
In this note, we consider the decomposition of unitary matrices (transformations)
into simple unitary matrices with special structural requirement.

Recall that a two-level $d\times d$ unitary matrix is a unitary matrix obtained from the
$d\times d$ identity matrix $I_d$ by changing a $2\times 2$ principal submatrix.
It is well known that every $d\times d$ unitary matrix can be decomposed into the product
of no more than $d(d-1)/2$ two-level unitary matrices. Among many applications,
this result has important implications to quantum computation.

We  will present a general scheme to decompose a $d$-by-$d$ unitary matrix
as the product of two-level unitary matrices with additional structure
and prescribed determinants. In particular, the decomposition can be
done by using two-level matrices in $d-1$ classes, where each class is
isomorphic to the group of $2\times 2$ unitary matrices. The proposed scheme
is easy to apply, and useful in treating problems with additional
structural restrictions.   In particular, the result will be used to deduce
the result in \cite{Vet}
that every quantum (unitary) gate acting on $n$-qubit register can be
expressed as no more than $2^{n-1}(2^n-1)$ fully controlled single-qubit
gates chosen from $2^n-1$ classes so that the quantum gates in each class
share the same $n-1$ control qubits.
 Moreover, it is shown that 
it is easy to adjust our  decomposition scheme
to take advantage of additional structure evolving in the process.

In Section 2, we will present the proposed decomposition scheme, and describe
some basic applications.
A Matlab program is written based on the proposed scheme, and is available at
http://cklixx.people.wm.edu/mathlib.html.
Some notes about the program will be given at the end of Section 2.
In Section 3, we describe the implication of our result to quantum computation.
a short conclusion will be given in Section 4.

\section{Basic results and examples}

Let  $P = (j_1, j_2, \dots, j_d)$ be such that the entries of $P$
correspond to a permutation of $(1, 2, \dots, d)$. A two-level
unitary matrix is called a $P$-unitary matrix of type $k$ for $k
\in \{1, 2, \dots, d-1\}$ if it is obtained from $I_d$ by changing
a principal submatrix with row and column indexes $j_k$ and
$j_{k+1}$. For example, if $P=(j_1, j_2, j_3, j_4) = (1,2,4,3)$,
then the three types of $P$-unitary matrices of type $1$, $2$,
and $3$ have the forms
$$
\pmatrix{
* & * & 0 & 0 \cr
* & * & 0 & 0 \cr
0 & 0 & 1 & 0 \cr 0 & 0 & 0 & 1 \cr}, \quad \pmatrix{ 1 & 0 & 0 &
0 \cr 0 & * & 0 & * \cr 0 & 0 & 1 & 0 \cr 0 & * & 0 & * \cr},
\quad \rm {\mbox and}\quad\pmatrix{ 1 & 0 & 0 & 0 \cr 0 & 1 & 0 &
0 \cr 0 & 0 &
* &
* \cr 0 & 0 & * & * \cr},$$  respectively .
We have the following.

\begin{proposition} \label{2.1} Every
$d$-by-$d$ unitary matrix $U$ can be written as a product of no more
than $d(d-1)/2$ $P$-unitary matrices. Moreover, these $P$-unitary
matrices can be chosen to have any determinants with modulus 1 as
long as their product equals $\det(U)$.
\end{proposition}

An immediate consequence of the proposition is the following.

\begin{corollary}  Every $d\times d$ special unitary matrix
can be written as a product of no more than $d(d-1)/2$
$P$-unitary matrices with
determinant 1. \rm
\end{corollary}

It is  instructive to illustrate a special case of the proposition. We consider the
case when $d = 4$ and $P = (j_1,j_2,j_3,j_4) = (1,2,4,3)$ as above.
(As we will see, this example is
relevant to the discussion on quantum gates in Section 3.)

Let
$$ U=\pmatrix{
a_{11} & a_{12} & a_{13} & a_{14} \cr a_{21} & a_{22} & a_{23} &
a_{24} \cr a_{31} & a_{32} & a_{33} & a_{34} \cr a_{41} & a_{42} &
a_{43} & a_{44}\cr}$$ be a four-by-four unitary matrix. Let
$\mu_1,\mu_2,...,\mu_6$ be such that $\mu_1,\mu_2,...,\mu_6\in
\{z: |z|=1\}$ and $\mu_1\mu_2\cdots \mu_6=\det(U)$. We divide the
construction into two steps.

\medskip\noindent
{\bf Step 1.} We consider the column of $U$ labeled by the first
entry of $P$ (i.e., the first column).

Choose $P$-unitary matrix $U_1$ of type $3$ with
$\det(U_1)=\bar{\mu}_1$ such that the $(j_4,j_1)=(3,1)$ entry of
$U_{1}U$ is 0 as follows. Let $u_1=\sqrt{|a_{31}|^2+|a_{41}|^2}$
and
$$  U_{1}= \pmatrix{1 &0 & 0 & 0 \cr 0& 1 & 0 & 0 \cr 0
& 0 & \frac{\bar{\mu}_1a_{41}}{u_1} &
\frac{-\bar{\mu}_1a_{31}}{u_1} \cr 0 & 0&\frac{\bar{a}_{31}}{u_1}
& \frac{\bar{a}_{41}}{u_1}\cr}. \quad \mbox
{\rm Then}\quad
U_1U=
\pmatrix{a_{11} & a_{12} & a_{13} & a_{14} \cr
a_{21} & a_{22} & a_{23} & a_{24} \cr
0 & a'_{32} & a'_{33} & a'_{34} \cr
u_1 & a'_{42} & a'_{43} & a'_{44} \cr}.$$

Next choose $P$-unitary matrix $U_{2}$ of type $2$ such that
$(j_3,j_1)=(4,1)$ entry of $U_{2}U_{1}U$ is 0 as follows. Let
$u_2=\sqrt{|a_{21}|^2+u_1^2}$, and
$$U_{2}=\pmatrix{1& 0 & 0 & 0
\cr 0 & \frac{\bar{a}_{21}}{u_2} & 0 & \frac{u_1}{u_2} \cr 0 & 0 &
1 & 0 \cr 0& \frac{-\bar{\mu}_2u_1}{u_2} & 0 &
\frac{\bar{\mu}_2a_{21}}{u_2}\cr}. \quad {\rm Then} \quad
U_{2}U_{1}U=\pmatrix{a_{11} & a_{12} & a_{13} & a_{14} \cr u_2 &
a'_{22} & a'_{23} & a'_{24} \cr 0 & a'_{32} & a'_{33} & a'_{34}
\cr 0 & a''_{42} & a''_{43} & a''_{44} \cr}.$$

Now, choose $P$-unitary matrix $U_{3}$ of type $1$ so that the
$(j_2,j_1) = (2,1)$ entry of $U_{3}U_{2}U_{1}U$ is also 0, and the
$(1,1)$ entry of $U_{3}U_{2}U_{1}U$ equals 1 as follows. Let
$$U_{3}={\small \pmatrix{\bar{a}_{11}& u_2 &
0&0 \cr -\bar{\mu}_3u_2 & \bar{\mu}_3a_{11} & 0 &0 \cr 0 & 0 &
1&0\cr 0&0&0&1 \cr}}.\quad  {\rm Then}\quad
V=U_{3}U_{2}U_{1}U={\small \pmatrix{1 & 0 & 0 & 0 \cr 0 & a''_{22} &
a''_{23} & a''_{24} \cr 0 & a'_{32} & a'_{33} & a'_{34} \cr 0 &
a''_{42} & a''_{43} & a''_{44} \cr}}.$$
Note that the first row of
$V$ has the form $(1,0,0,0)$ because $V$ is unitary.

\medskip \noindent
{\bf Step 2.}  We turn to columns of $V$ labeled by $j_2=2$ and
$j_3=4$.

Choose $P$-unitary matrices $U_{4}, U_5$ of types 3 and 2 with
determinants $\bar{\mu}_4$ and $\bar{\mu}_5$, respectively, so
that the $(j_4,j_2) = (3,2)$ entry of $U_4V$ and the $(j_3,j_2) =
(4,2)$ entry of $U_5U_4V$ are 0. Then choose a $P$-unitary matrix
$U_6$ of type $3$ with determinant $\bar{\mu}_6$ so that the
$(j_4,j_3) = (3,4)$ entry of $U_6U_5U_4V$ is 0. Here are the zero
patterns of the matrices in the process:
$$U_4V = {\small\pmatrix{1 & 0 & 0 & 0 \cr 0 & * & * & * \cr
0 & 0 & * & * \cr 0 & * & * & * \cr}},
\ U_5U_4V = {\small \pmatrix{1 & 0 & 0 & 0 \cr 0 & 1 & 0 & 0 \cr 0 & 0 & *
& * \cr 0 & 0 & * & * \cr}}, \ U_6U_5U_4V = {\small \pmatrix{1 & 0 & 0 & 0
\cr 0 & 1 & 0 & 0 \cr 0 & 0 & 1 & 0 \cr 0 & 0 & 0 & 1 \cr}}.$$
 Note
that in the last step, the $(j_4, j_4)=(3,3)$ entry of
$U_6U_5U_4V$ is 1 because
$$\det(U_6U_5U_4U_3U_2U_1)\det(U)=\bar{\mu}_6\bar{\mu}_5\cdot\cdot\cdot\bar{\mu}_1\det(U)
=|\det(U)|^2=1.$$
Consequently,
$$U=U_1^{\dag}U_2^{\dag}U_3^{\dag}U_4^{\dag}U_5^{\dag}U_6^{\dag}.$$
Clearly, each $U_j^{\dag}$ is a $P$-unitary matrix of the same
type as $U_j, j=1,2,...,6.$

Obviously, we can skip some of the $P$-unitary matrices if the
entry to be eliminated is already 0 during the process. We can now
present the proof of Proposition 2.1.

\medskip\noindent
{\bf Proof of Proposition \ref{2.1}.} \rm Let $P = (j_1, j_2, \dots,
j_d)$ where the entries of $P$ are a permutation of $(1, 2,\dots,
d)$. Let $U$ be a $d$-by-$d$ unitary matrix. We extend the
construction in the example to the general case as follows.

\smallskip
{\bf Step 1.} First consider the $j_1$th column of $U$. One can
choose $P$-unitary matrices $U_{d-1}, U_{d-2}, \dots$, $U_1$ of
types ${d-1}, {d-2}, \dots, 1$ with prescribed determinants
to eliminate the entries of $U$ in positions $(j_d, j_1),
(j_{d-1},j_1), \dots, (j_2,j_1)$ successively, so that the $j_1$th
column of the matrix $V = U_1 U_2 \cdots U_{d-1}U$ equals the
$j_1$th column of $I_d$. Since $V$ is unitary, the $j_1$th row of
$V$ will equal the $j_1$th row of $I_d$.

\smallskip
{\bf Step 2.} Consider the $j_2$th column of $V$. One can choose
$P$-unitary matrices $V_{d-1}, V_{d-2}, \dots, V_2$ of types
${d-1}, {d-2}, \dots, 2$ with prescribed determinants to
eliminate the entries of $V$ in positions $(j_d, j_2),
(j_{d-1},j_2), \dots, (j_3,j_2)$ successively, so that the $j_2$th
column of the matrix $W = V_2 \cdots V_{d-1}V$ equals the $j_2$th
column of $I_d$. Since $W$ is unitary, the $j_2$th row of $W$ will
equal the $j_2$th row of $I_d$.

\smallskip
We can repeat this process in converting the $j_3$th $\dots,
j_{d-1}$th columns to the $j_3$th $\dots, j_{d-1}$th column of
$I_d$ successively, by $P$-unitary matrices. Finally, the
$(j_d,j_d)$ element will also be 1 by the determinant condition
imposed on the $P$-unitary matrices. Multiplying the inverses of
the $P$-unitary matrices in the appropriate orders, we see that
$U$ is a product of $P$-unitary matrices as asserted, and the
number of $P$-unitary matrices used is no more than $(d-1) +
\cdots + 1 = d(d-1)/2$ because some of the $P$-unitary matrices
may be chosen to be identity if the entry to be eliminated is
already 0 during the process. \qed

\medskip
Several remarks are in order.
First, if we choose $P = (1,\dots, d)$, then $P$-unitary matrices
are tridiagonal unitary matrices. For example, using
$P = (1,2,3,4)$, we can decompose a $4\times 4$ unitary $U$ as
{\footnotesize
$$U =
\pmatrix{1 & & & \cr & 1 & & \cr & & * & * \cr & & * & * \cr}
\pmatrix{1 & & & \cr & * & * & \cr & * & * & \cr & & & 1\cr}
\pmatrix{* & * & & \cr * & * & & \cr & & 1 & \cr & & & 1 \cr}
\pmatrix{1 & & & \cr & 1 & & \cr & & * & * \cr & & * & * \cr}
\pmatrix{1 & & & \cr & * & * & \cr & * & * & \cr & & & 1\cr}
\pmatrix{1 & & & \cr & 1 & & \cr & & * & * \cr & & * & * \cr}.$$}

Second, we can use our result to deduce some classical results. For example,
the same result holds for real orthogonal matrices. In particular,
every special orthogonal matrix $U$ can be decomposed as a product of two-level
special orthogonal matrices. For instance, a classical result
asserts that every $3\times 3$ special orthogonal matrix
$U$, which is known as a rotation,
can be written as the product of special orthogonal matrices of the form
{\small $$U = \pmatrix{1 & & \cr & * & * \cr & * & * \cr}
\pmatrix{* & * & \cr  * & * & \cr &  & 1 \cr}
\pmatrix{1 & & \cr & * & * \cr & * & * \cr}.$$}

\noindent
Geometrically, it means that every rotation $U$ in $\IR^3$ can be achieved by
a rotation of the $yz$-plane, and then a rotation of the $xy$-plane, and followed by
a rotation of the $yz$-plane again. Note that one can also express $U$ as the
product of special orthogonal matrices of the form
{\small $$U = U_1U_2U_3 = \pmatrix{* & * & \cr  * & * & \cr &  & 1 \cr}
\pmatrix{* & & *\cr & 1 &  \cr * &  & * \cr}
\pmatrix{1 & & \cr & * & * \cr & * & * \cr}$$}
as suggested in \cite[p.26]{H}.
Here, we need only to find $U_1, U_2, U_3$ in the above form
so that $U_1^{\dag}U$ has zero $(2,1)$ entry,
$U_2^{\dag}U_1^{\dag}U$ has $(3,1)$ entry also equal to zero,
and $U_3^{\dag}U_2^{\dag}U_1^{\dag}U$
has $(3,2)$ entry also equal to zero.
Thus, we can express $U$ as a rotation of the $yz$-plane, and then a rotation of
the $xz$-plane, and followed by a rotation of the $xy$-plane.
This illustrates the flexibility of our decomposition scheme.

Furthermore, our scheme is easy to implement. A Matlab program {\tt pud.m} is written
and available at http://cklixx.people.wm.edu/mathlib.html.
Applying the Matlab command
$$A = {\tt pud}(U,[i_1,\dots,i_n])$$
to an $n\times n$ unitary $U$ and a specific permutation $P = (i_1,\dots, i_n)$
will yield  $P$-unitary matrices $A\{1\}, \dots, A\{N\}$ with $N = n(n-1)/2$
such that $A\{1\}\cdots A\{N\}U=I_N$.

\section{Decomposition of Quantum Gates}

We will apply Proposition \ref{2.1} to the decomposition of quantum gates.
Here are some basic background.
The foundation of quantum computation \cite{NC} involves
the encoding of computational tasks into the temporal
evolution of a quantum system. A register of
qubits, identical two-state quantum systems, is employed, and
quantum algorithms can be described by unitary
transformations and projective measurements acting on
the $2^n$-dimensional state vector of the register. In this
context, unitary matrices (transformations) of size $2^n$
are called quantum gates.

It is known \cite{Bet} ( see also \cite[\S 4.6]{NO} and
\cite[pp.188-193]{NC}) that the set of
single-qubit gates (acting on one of the $n$ qubits)
and CNOT gates (acting on two of the $n$ qubits)
are universal.  In other words, every unitary gate acting on
$n$-qubit register can be implemented with single-qubit gates
and CNOT gates.
The conventional approach of reducing an arbitrary $n$-qubit gate
into elementary gates is done as follows.

\medskip\noindent
{\bf Step 1.} Decompose a $2^n\times 2^n$ general unitary gate
into the product of two-level unitary matrices and find
a sequence of ${\rm C}^{n-1}V$ and ${\rm C}^{n-1}$NOT gates which
implements each of them.

\medskip\noindent
{\bf Step 2.} Decompose
${\rm C}^{n-1}V$ and ${\rm C}^{n-1}$NOT gates into
single qubit gates and CNOT gates, which are referred to as elementary gates,
for physical implementation.

\medskip
Note that a fully controlled qubit gate ${\rm C}^{n-1}V$ has $(n-1)$ control qubits,
each of which has the value 0 or 1, specify the subspace
in which that gate $V = \pmatrix{v_{11}& v_{12}\cr v_{21}& v_{22}\cr}$
operates. When $V = {\scriptstyle \pmatrix{ 0 & 1 \cr 1 & 0 \cr}}$, 
the ${\rm C}^{n-1}V$ gate reduces to the ${\rm C}^{n-1}$NOT gate.
For example, for a two-qubit system, quantum gates
are $4\times 4$ unitary matrices with rows and columns labeled by
the binary sequences $00, 01, 10, 11$.
If we use the first qubit to control the second qubit, then the two
controlled single-qubit gates are:
{\small $$
\begin{array}{c} \cr 00\ \cr 01\ \cr 10\ \cr 11\ \cr\end{array} \hskip -.11in
\begin{array}{c} ~~ 00 ~~\ 01 ~~\ 10 ~~\ 11 ~~\cr
\pmatrix{
v_{11} & v_{12} & 0 & 0 \cr
v_{21} & v_{22} & 0 & 0 \cr
0 & 0 & 1 & 0 \cr
0 & 0 & 0 & 1 \cr}
\end{array}, \quad
\begin{array}{c} \cr 00\cr 01\cr 10\cr 11 \cr\end{array} \hskip -.11in
\begin{array}{c} \, 00 ~~ \ 01 ~~ \ 10 ~~ \ 11 ~~ \cr
\pmatrix{
1 & 0 & 0 & 0 \cr
0 & 1 & 0 & 0 \cr
0 & 0 & v_{11} & v_{12} \cr
0 & 0 & v_{21} & v_{22} \cr}
\end{array}.$$}
If we use the second qubit to control the first qubit, then the two
controlled single qubit gates are:
{\small $$
\begin{array}{c} \cr 00~~~ \ \cr 01~~~ \ \cr 10~~~ \ \cr 11~~~ \ \cr\end{array} \hskip -.3in
\begin{array}{c} 00 ~~\ 01~~ \ 10~~ \ 11 \cr
\pmatrix{
v_{11} & 0 & v_{12} & 0 \cr
0 & 1 & 0 & 0 \cr
v_{21} & 0 & v_{22} & 0 \cr
0 & 0 & 0 & 1 \cr}
\end{array}, \quad
\begin{array}{c} \cr 00  \cr 01  \cr 10  \cr 11  \cr\end{array} \hskip -.1in
\begin{array}{c} ~~00~~ \ 01~~ \ 10~~ \ 11~~ \cr
\pmatrix{
1 & 0 & 0 & 0 \cr
0 & v_{11} & 0 & v_{12} \cr
0 & 0 & 1 & 0 \cr
0 & v_{21} & 0 & v_{22} \cr}
\end{array}.$$}

\noindent
In general, if we label the rows and columns of quantum gates (unitary matrices) acting on
$n$ qubits by binary sequences $x_1\cdots x_n$, then a ${\rm C}^{n-1}V$ gate corresponds
to a two-level matrix obtained from $I_{2^n}$ by replacing its $2\times 2$ principal submatrix
lying in rows and columns $X = x_1 \cdots x_n$ and $\tilde X  = \tilde x_1 \dots \tilde x_n$
by $V$ for two binary sequence $X$ and $\tilde X$ differ exactly in one of their terms, say,
$x_i\ne \tilde x_i$.

Note that if $P = (1,2,4,3)$ as in the beginning of Section 2, then the $P$-unitary
matrices are ${\rm C}^1$V gates acting on 2-qubit register. By Proposition 2.1,
every quantum gate on 2-qubit register is a product of ${\rm C}^1$V gates and no
${\rm C}^1$NOT gates needed.
We can extend this conclusion to $n$-qubit quantum gates as follows.

Assume that $n$ is a
positive integer and $N = 2^n$. A Gray code $G_n$
\cite{S} is an $N$-tuple $G_n = (X_1, X_2, \dots, X_N)$
such that

\begin{itemize}
\item[(a)] each $X_1, X_2, \dots, X_N$ are length $n$ binary
sequences corresponding to binary representation of the numbers
$0, 1, \dots, N-1$, arranged in a certain order,

\item[(b)] two adjacent sequences $X_j$ and $X_{j+1}$ differ in
only one position for each $j = 1, 2, \dots, N-1$,

\item[(c)] the sequences $X_N$ and $X_1$ differ in only one position.
\end{itemize}

One can construct a Gray code $G_n$ recursively as follows.

\medskip\noindent \it
Set $G_1 = (0,1)$; for $n \ge 1$ and $N = 2^n$, if  $G_n = (X_1,
\dots, X_{N-1}, X_{N})$, let
$$G_{n+1} = (0X_1, \dots, 0X_{N-1}, 0X_N, 1X_N, 1X_{N-1}, \dots, 1X_1).$$
\rm

\smallskip
\noindent For example, we have
$G_2 = (00, 01, 11, 10), \quad G_3 = (000,001,011,010,110,111,101,100),  \hbox{ etc.}$

\smallskip
One easily adapts the definition of $P$-unitary matrices to define
$G_n$-{\it unitary matrices}, which  correspond to controlled
single-qubit gates in quantum information science. To this end,
label the rows and columns of an $N\times N$ matrix by the binary
numbers $0\cdots 0, \ 0\dots01, \dots, \ 1\cdots 1.$
An $N\times N$ two-level unitary matrix is a $G_n$-unitary matrix
of type $k$ if it differs from $I_N$ by a principal submatrix
with rows and columns labeled by two consecutive terms $X_k$ and
$X_{k+1}$ in the Gray code $G_{n}=(X_{1},X_{2},...,X_{N})$, $k \in
\{1, 2, \dots, N-1\}$.
Clearly, there are
$N-1$ types of $G_n$-unitary matrices. Since $X_k$ and $X_{k+1}$
differ in only one position, every $G_n$-unitary matrix
corresponds to a ${\rm C}^{n-1}$V gate.
It is now easy to adapt Proposition \ref{2.1} to prove the following.

\begin{proposition} \label{main} Let $n$ be a positive integer and $N = 2^n$.
Every $N\times N$ unitary matrix $U$ is a product of $m$
$G_n$-unitary matrices with $m\leq N(N-1)/2$. Furthermore, the
$G_n$-unitary matrices can be chosen to have any determinant with
modulus 1 as long as their product equals  $\det(U)$.
\end{proposition}

\it Proof. \rm Identify $(0, 1,\dots, N-1)$ with
the $N$-tuple of binary numbers $(0\cdots0, 0\cdots01, \ \dots, \ 1\cdots 1)$;
label the rows and columns of $U$ by the binary numbers
$0\cdots0, 0\cdots01, \ \dots, \ 1\cdots 1$. Then apply Proposition \ref{2.1}
to $U$ with $P$ replaced by $G_n$. \qed

To illustrate Proposition \ref{main}, consider a  quantum gate $U$ acting on 3 qubits.
Label its rows and columns by $000,\dots,111$, and consider the Gray code
$G_3 = (000,001,011,010,110,111,101,100).$
Then the $G_3$ permutation sequence corresponds to
$P = (1,2,4,3,7,8,6,5)$. One may find $G_3$-matrices
$U_1, \dots, U_{28}$ to create zero entries in the following order:

\smallskip
column 1: create zeros at the $(5,1), (6,1), (8,1), (7,1), (3,1), (4,1), (2,1)$ positions;

column 2: create zeros at the $(5,2), (6,2), (8,2), (7,2), (3,2), (4,2)$ positions;

column 4: create zeros at the $(5,4), (6,4), (8,4), (7,4), (3,4)$ positions;

column 3: create zeros at the $(5,3), (6,3), (8,3), (7,3)$ positions;

column 7: create zeros at the $(5,7), (6,7), (8,7)$ positions;

column 8:  create zeros at the $(5,8), (6,8)$ positions;

column 6: create zeros at the $(5,6)$ position.

\noindent
Note we deal with the columns in the order of $1,2,4,3,7,8,6$, and eliminate the entries
in each column in the order of $5,6,8,7,3,4,2,1$ if it is not yet zero.

\medskip
Proposition \ref{main} helps us to decompose a quantum gate acting on
$n$ qubits as the product of no more than
$2^{n-1}(2^n-1)$ fully controlled single-qubit gates. One can then apply the techniques in \cite{Bet}
or \cite{Vet} to further decompose them into elementary gates, i.e., single qubit gates
and CNOT gates.
In fact, the authors in \cite{Vet} use the Gray code techniques
to relabel the computational basis and then use a clever scheme to reduce
the number of control bits. It led to a very efficient decomposition
of quantum gates. For other papers concerning efficient and
practical algorithms in constructing unitary gates; see
\cite{Det,Di,K,M1,M2,NO,NC,Vet} and the references therein.
Note that
relabeling the computational basis using the Gray code is
basically the same as our procedure. In our decomposition scheme, we focus on
the order of creating zero entries in the process. It is easy to implement,
and easy to change if new structure and (zero) patterns evolve in the process.
For instance, in the 3-qubit case, if the unitary $U$ is given as or becomes 
$I_{2}\bigoplus V\bigoplus I_1$ after a few steps, then we can focus our P-unitary 
matrices with rows and columns associated with the subsequences
$(011, 010, 110, 100, 101)$ corresponding to $(4,3,7,5,6)$. Then we can use this smaller class of P-unitary matrices to eliminate all the non-zero entries in off-diagonal positions
for the rest of the process.

\section{Discussion}

In this note, we obtained a decomposition of a $d$-by-$d$ unitary
matrix as product of special structures specified by a vector $P =
(j_1, j_2, \dots, j_d)$ such that the entries of $P$ correspond to
a permutation of $(1,2,\dots, d)$. The result is then applied to
show that every unitary gate $U$ acting on $n$ qubits can be
decomposed as product of special two-level unitary matrices
corresponding to fully controlled single-qubit gates. This was done by
using Gray code $G_n = (X_1, X_2, \dots, X_N)$ with $N = 2^n$, and
constructing $G_n$-unitary matrices which are two-level matrices
obtained from $I_N$ by changing its principal submatrix with row
and column indexes $X_k$ and $X_{k+1}$ for $k = 1, 2, \dots, N-1$.

\smallskip
There are other applications of Proposition \ref{2.1}. For example, if
$P = (1, 2, \dots, d)$, then $P$-unitary matrices are two-level
tridiagonal unitary matrices. In numerical linear algebra and
other applications,  it is useful to decompose a matrix into
tridiagonal forms with simple structure; e.g., see \cite{F,St} and
their references. In the context of
quantum information science, 
it is desirable to have a decomposition scheme of the unitary matrix
respecting the tensor (Kronecker) product structure. The
same technique might be useful for decomposition of matrices with
other multilinear structures. Moreover,  
it is easy to adjust our decomposition scheme
to take advantage of additional structure evolving in the process.

\medskip
\noindent
{\bf Acknowledgment}

\smallskip
The authors would like to thank the referees for their helpful comments
leading to the improvement of the paper.
They also thank Yiu-Tung Poon for drawing our attention to the references \cite{M1,M2}.

Li was  supported by a USA NSF grant, and a HKU RGC grant; he was an
honorary professor of Taiyuan University of Technology (100 Talent Program scholar),
an honorary professor of the University of Hong Kong, and
an honorary professor of Shanghai University.
Roberts was supported by an NSF CSUMS grant.
Yin was supported by National Science Foundation of China
(11101322) and the Fundamental Research Funds for the Central
Universities in Xidian University of 2013. The research was done while she was
visiting the College of William and Mary during the academic year
2012-13 under the support of China Scholarship Council.

\end{document}